\documentclass[letter,traditabstract]{aa}  

\usepackage{graphicx}
\usepackage{txfonts}
\newcommand{\msun}{\mbox{$M_{\odot}$}}
\newcommand{\Msun}{\mbox{$M_{\odot}$}}
\newcommand{\lsun}{\mbox{$L_{\odot}$}}
\newcommand{\Lsun}{\mbox{$L_{\odot}$}}

\newcommand{\Zsun}{\mbox{$Z_{\odot}$}}
\newcommand{\zsun}{\mbox{$Z_{\odot}$}}
\newcommand{\teff}{\mbox{$T_{\rm eff}$}}

\newcommand{\mdot}{\mbox{$\dot{M}$}}

\begin{document}

   \title{The maximum black hole mass at solar metallicity}

   \author{Jorick S. Vink, Gautham N. Sabhahit, Erin R. Higgins \inst{1}}

   \institute{Armagh Observatory and Planetarium, College Hill,
              BT61 9DG Armagh, Northern Ireland\\
              \email{jorick.vink@armagh.ac.uk}
             }

   \date{Received 8 May 2024, Accepted 10 July 2024}

% \abstract{}{}{}{}{} 
% 5 {} token are mandatory
 
  \abstract
{We analyse the current knowledge and uncertainties in detailed stellar evolution and wind modelling to evaluate the mass of the most massive stellar black hole (BH) at solar metallicity. Contrary to common expectations that it is the most massive stars that produce the most massive BHs, we find that the maximum $M_{\rm BH}^{\rm Max} \simeq 30 \pm 10$ \msun\ is found in the canonical intermediate range between $M_{\rm ZAMS}$ $\simeq$ 30 and 50\msun\ instead. The prime reason for this seemingly counter-intuitive finding is that very massive stars (VMS) have increasingly high mass-loss rates that lead to substantial mass evaporation before they expire as stars and end as lighter BHs than their canonical O-star counterparts.}


\maketitle
%
%-------------------------------------------------------------------

\section{Introduction}

Recently, a black hole (BH) of 33\,\msun\ was discovered in the Galaxy by GAIA \citep{GAIABH3}. The mass of GAIA BH3 is substantially higher than the heaviest Population (Pop) {\sc I} BHs, which are 10-20\,\msun\ \citep{belc10,Miller-jones}. The most logical explanation for the high BH mass in GAIA BH3 is a low metallicity ($Z$). 
Low $Z$ was also invoked for the first BHs that were detected with gravitational waves \citep{abbott16}, as expected from $Z_{\rm Fe}$-dependent winds of Wolf-Rayet (WR) stars \citep{vdek05}, which naturally produce lower BH masses for solar\footnote{We use solar metallicity ($\Zsun$) when referring to Pop {\sc I} stars with Galactic metallicity, as the Milky Way also contains stars and BHs that formed at earlier times, when $Z$ was lower.} $Z$ \citep{EV06}. 

It was therefore a surprise that LB-1 was announced to involve a Pop I 70\msun\ BH \citep{liu19}, and while the observational arguments have mostly dissipated \citep{abdul20,el-baldry20}, the key question of whether it would be theoretically possible to produce such a heavy BH at solar $Z$ is fundamental. In this Letter,
we analyse the available knowledge and uncertainties on both the internal mixing in massive stars and their stellar wind mass loss to address the question of the maximum BH mass at $\Zsun$. This is particularly relevant as there appears to be a misconception that the most massive stars produce the heaviest BHs.  

Over the last 5 years, we have made a dedicated effort to systematically study the key ingredients relevant for massive star evolution, which vary because of uncertainties in core-boundary mixing (CBM) \citep{higgins19} and wind mass loss \citep{Sabh22}. Mass loss is particularly influential for the most massive stars, where stars over 100\msun\ are referred to as very massive stars (VMS), as well as other mixing processes, including semi-convection or additional energy transfer in super-adiabatic layers.  

As there are many large uncertainties in interior mixing and wind mass loss (see Sect.\,2), it is not possible to accurately predict the final stellar or BH mass for a given star of a given initial mass. However, using the 
relevant observational and theoretical constraints available, we show that it is feasible to compartmentalise these issues, and to provide a realistic maximum BH mass. 

We previously performed the analysis for the maximum BH mass at {\it low} $Z$, where we used our knowledge of interior mixing and wind mass loss and obtained a blue supergiant (BSG) progenitor for an 85\msun\ BH, such as was detected in GW\,190521 \citep{Vink21,Winch24}.
Our studies lead to a maximum BH mass as a function of cosmic time, where at low $Z$ (below $\sim$10\% $\Zsun$), the maximum BH mass is set by pair-instability physics \citep{woosley17,farmer19,marchant19,Renzo2020}, but at high $Z$, where the maximum BH mass is anticipated to be set by wind mass loss, this number still needs to be determined. This is the purpose of this Letter. 

In \cite{Higg21} we investigated the BH masses during the classical WR stage, but we did not study the evolution into this regime, which we undertook during complementary investigations \citep{higgins19,Sabh22}. In this Letter, we use our overall knowledge of interior mixing, wind mass loss, and the physics and location of the upper luminosity limit of red supergiants (RSGs) as part of the Humphreys-Davidson (HD) limit \citep{Hump79}, to indirectly accommodate for eruptive mass loss to infer the maximum BH mass at $\Zsun$.

\section{Stellar evolution models}

We computed a detailed set of MESA \citep{paxton2011,paxton2013,paxton2015,paxton2018,paxton19} stellar evolution models for stars up to 250\,\Msun\ using the exact same mass-loss implementation and additional physical inputs as for the core H-burning models from \cite{Sabh22}. For the main sequence, we employed the canonical \cite{vink2001} wind mass-loss recipe, 
but above the transition mass-loss rate, we employed the high $\Gamma$ wind relation of \cite{vink2011}. 
These main-sequence models do not require the usage of MLT$++$ during core H-burning due to the wind enhancement. Beyond core H-burning, we did employ MLT$++$ with the same parameters as in \cite{Sabh21}, which was calibrated to the upper RSG luminosity of the HD limit of $\log{L/\lsun} = 5.5$ \citep{Davies20,Macdonald22}. Following \cite{Vink21}, we switched 
to the RSG recipe of \cite{deJager1988} below 4 kK, while we adopted the new hydrodynamically consistent WR recipe of \cite{sv20} above 100 kK.

 \subsection{Uncertainties regarding interior mixing}

Massive stars have convective cores but radiative envelopes. The size of the core can be extended by CBM \citep{Anders23}, also known as overshoot.  Higher amounts of core overshoot extend the main-sequence lifetime, and thereby the total amount of mass lost in winds.

Until 2010, most modellers employed no overshoot, or relatively small amounts of CBM parametrised in terms of a step overshooting value $\alpha_{\rm ov} \simeq 0-0.1$ in detailed stellar evolution models \citep{meynet00,heger00}. However, in \cite{vink10}, we suggested that given the unexpectedly high number of observed B supergiants, stellar models should be explored with $\alpha_{\rm ov}$ up to $\simeq$ 0.5, which widens the main sequence\footnote{Alternatively, the cooler objects could be stellar mergers \citep{Menon24}}. 

Ever since, we have taken the pragmatic approach to explore both low values of $\alpha_{\rm ov} = 0.1$ and higher values of $\alpha_{\rm ov} = 0.5$ in our canonical stellar models to accommodate for the full range of mixing uncertainties revealed by asteroseismology \cite{bowman2020}\footnote{So far, this is only feasible for B-type stars up to $\simeq 20 \msun$.} and eclipsing binaries\footnote{which are possible at higher masses into the regime of O star, where mass loss becomes a relevant factor}. 

While recent advances in simulations of core convection (see \cite{lecoanet23} for a review) and \cite{Scott21} and \cite{herwig23} models indicate that more massive stars are expected to have larger amounts of CBM, asteroseismology studies of B stars indicate a roughly equal chance for low ($\alpha_{\rm ov} \leq 0.1$, \citet{Dupret04}) and high ($\alpha_{\rm ov} \simeq 0.44$; \citet{Briquet07}) values. 
The origin for what may be the root cause for this spread of CBM values for similar initial ZAMS masses remains to be revealed \citep{Costa19}.
For the meantime, we consider the equal empirical incidence of low and high  $\alpha_{\rm ov}$  for stars at a given initial mass as an instruction to consider that stars of equal masses could have a wide dispersion in their interior mixing properties.

It is challenging to perform similar asteroseismology investigations in the higher-mass regime because the effects of stronger stellar winds are more prevalent \citep{Aerts19}.  
Simply extrapolating the observational findings from asteroseismology for B-type stars into the more massive O-type regime is also questionable, but it is critical to know the $\alpha_{\rm ov }$ values because they set the core size, and thereby its explodability \citep{higgins19,temaj24}, as well as 
the maximum BH mass below pair instability \citep{Vink21,Tani21,Winch24}. 

From the use of eclipsing binaries in the ML-plane, $\alpha_{\rm ov }$ values \citep{higgins19} can potentially be obtained by disentangling the effects of interior mixing from wind mass loss. However, in the upper O-star mass regime of 70-100\msun\, the stellar HRD and ML-plane locations unfortunately become less sensitive to high and low $\alpha_{\rm ov }$ assumptions. 
Moreover, due to the larger convective cores, VMS become basically insensitive to the adopted value of $\alpha_{\rm ov}$, and the size of the relative pressure scale-height becomes smaller when moving up the mass range.

\subsection{Uncertainties in stellar winds}

Stellar winds of massive stars are thought to be driven by radiation pressure on spectral lines \citep{CAK}. 
It is important to realise that in the regimes with the highest uncertainties, the effects of wind mass loss on stellar mass evolution are generally weakest (e.g. in the weak-wind regime; see below). 

In domains in which the winds are most dominant, that is, in the regime of the VMS, they probably need to be boosted given both theoretical \citep{vink2011} and empirical evidence \citep{best14}. 
In \cite{Sabh22} we implemented boosted wind mass-loss above the \cite{vink2011} kink,

\begin{equation}
    \log \dot{M} \propto 4.77 \log(L/L_{\odot}) -3.99 \log(M_{\rm cur}/M_{\odot}).
\end{equation}
This enhanced mass-loss relation is switched above a specific wind efficiency number $\eta$, which sets the transition mass-loss rate. 
Alternative approaches have also been used in recent times. For instance, \cite{Romag24} employed a $\Gamma$-switch from an updated \cite{CAK} approach by \cite{best20}, but the value of $\Gamma$ on this {\it mathematical} transition has been shown to be insensitive to the CAK $\alpha$ parameter \citep{Sabh22}. Hence, we employed the original {\it physical} mass-loss transition point from \cite{vg12}, determined from the wind efficiency number $\eta$. 

We summarise our knowledge and uncertainties in wind mass loss over the full ZAMS mass range below.

\begin{itemize}

\item {\bf $M_{\rm ZAMS}$ $\la$20\msun}. In this mass range, stars are in the so-called {\it weak-wind regime} \citep{martins05}, and it is not yet established whether the cause is theoretical or diagnostic in nature. Nevertheless, even if stellar winds are weaker in this regime, it is important to realise that a huge uncertainty of 1 or 2 orders of magnitude is not expected to have a direct effect on the stellar mass evolution, and uncertainties in interior mixing \citep{Aerts19} are expected to dominate the stellar evolution.\\

\item {\bf 20$\msun$ $\la$ $M_{\rm ZAMS}$ $\la$ 60\msun}. In this regime, which we refer to as canonical massive stars, the uncertainties in mass-loss rates are substantial (by a factor of $\sim$3; \citep{krticka17,sundqvist19,vink22}), and as both mixing and mass loss play a role in the evolution of these stars \citep{langer12}, this regime is key for disentangling the separate effects \citep{higgins19}.\\

\item {\bf $M_{\rm ZAMS}$ $\ga$60\msun}. In this regime, the effect of winds dominates the evolution of interior mixing. While there could still be uncertainty at 60\msun, the uncertainty at the mass-loss transition point, which is at $\sim$80-100\msun\ for $\Zsun$, the mass-loss rates are the most accurate, within as little as 30\% as determined from the \cite{vg12} calibration study.\\

\end{itemize}

\subsection{Uncertainties in the treatment of radiative envelopes}

While we have standard ways of including convection via the mixing length theory (MLT), for objects close to the Eddington limit, such as in He-burning supergiants, radiative transport inside envelopes can become important, and structure models
predict strong inflation effects as well as density inversions to maintain hydrostatic equilibrium. Basically, one-dimensional (1D) treatment of radiative envelopes presents an unsolved problem in 1D stellar evolution modelling \citep{jiang15,Moens22a}.
In MESA, the envelopes can be treated in the standard MLT approach, or the convective
energy transport efficiency can be increased by employing MLT++ in super-adiabatic layers \citep{paxton2013,Sabh21}. In our analysis of the maximum BH mass, we considered both options. 

Moreover, it is not yet known whether episodic LBV-type mass-loss \citep{Grassitelli21} or eruptive mass loss \citep{Smith14,Cheng24} plays a role in addition to that by stellar winds discussed above. For this reason, we discuss the systematic uncertainties in envelope mixing and mass-loss physics for the 30-50\,\Msun\ range in terms of the HD limit in Sect.\,4.

\section{Results}

We present the results of our MESA evolution models over the entire ZAMS mass range up to 250\msun\ with the new mass-loss implementation as introduced in \cite{Sabh22}, and
we showcase a subset of these tracks in the HR diagram of Fig.\,1.
 We computed the stellar and wind evolution for the entire ZAMS mass range until the end of core He-burning. 
 
 \begin{figure}
   \centering
   \includegraphics[width=0.45\textwidth]{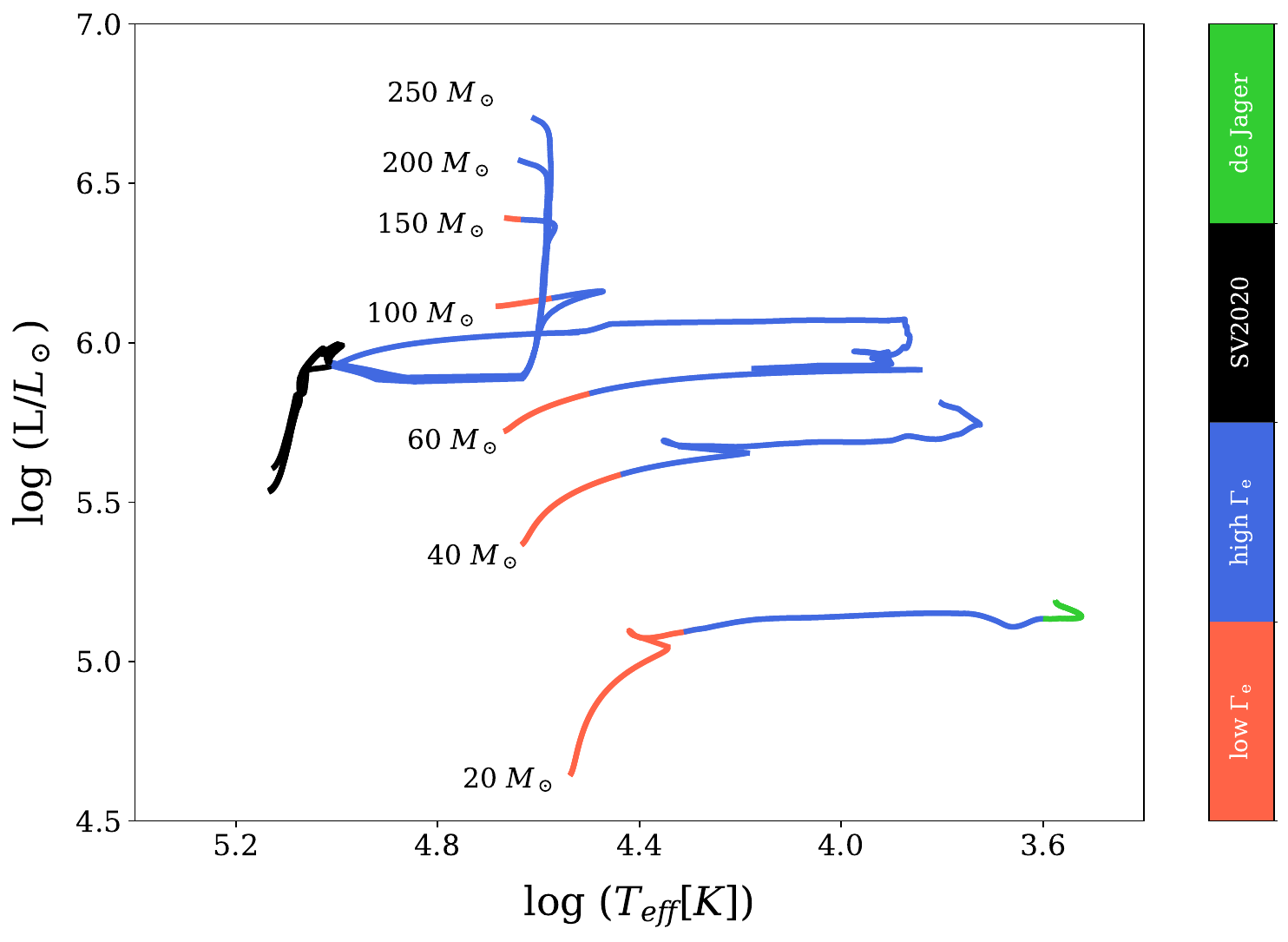}
   \caption{Hertzsprung-Russel diagram for MESA models for a range of initial masses using the mass-loss recipes given on the right-hand side. Stars above the transition point show vertical evolution, and lower-mass stars undergo classical horizontal redwards evolution. While the 20\,\msun\ model enters the RSG phase, the more massive models do not.}
        \label{fig:HRD}
    \end{figure}

    For a star of a given mass, we need to consider two basic aspects: (i) the duration ($\Delta t$) in a particular evolutionary state, for instance as a blue supergiant (BSG), RSG, or WR star (i.e. the characteristic $\teff$ of the evolutionary phase), and (ii) the mass-loss prescription in that \teff\ range: $\mdot(\teff)$, resulting in a total mass loss of

\begin{equation}
    \Delta{M} = \int \dot{M}(\teff) \Delta{t}. 
\end{equation}

\begin{figure}
   \centering
   \includegraphics[width=0.45\textwidth]{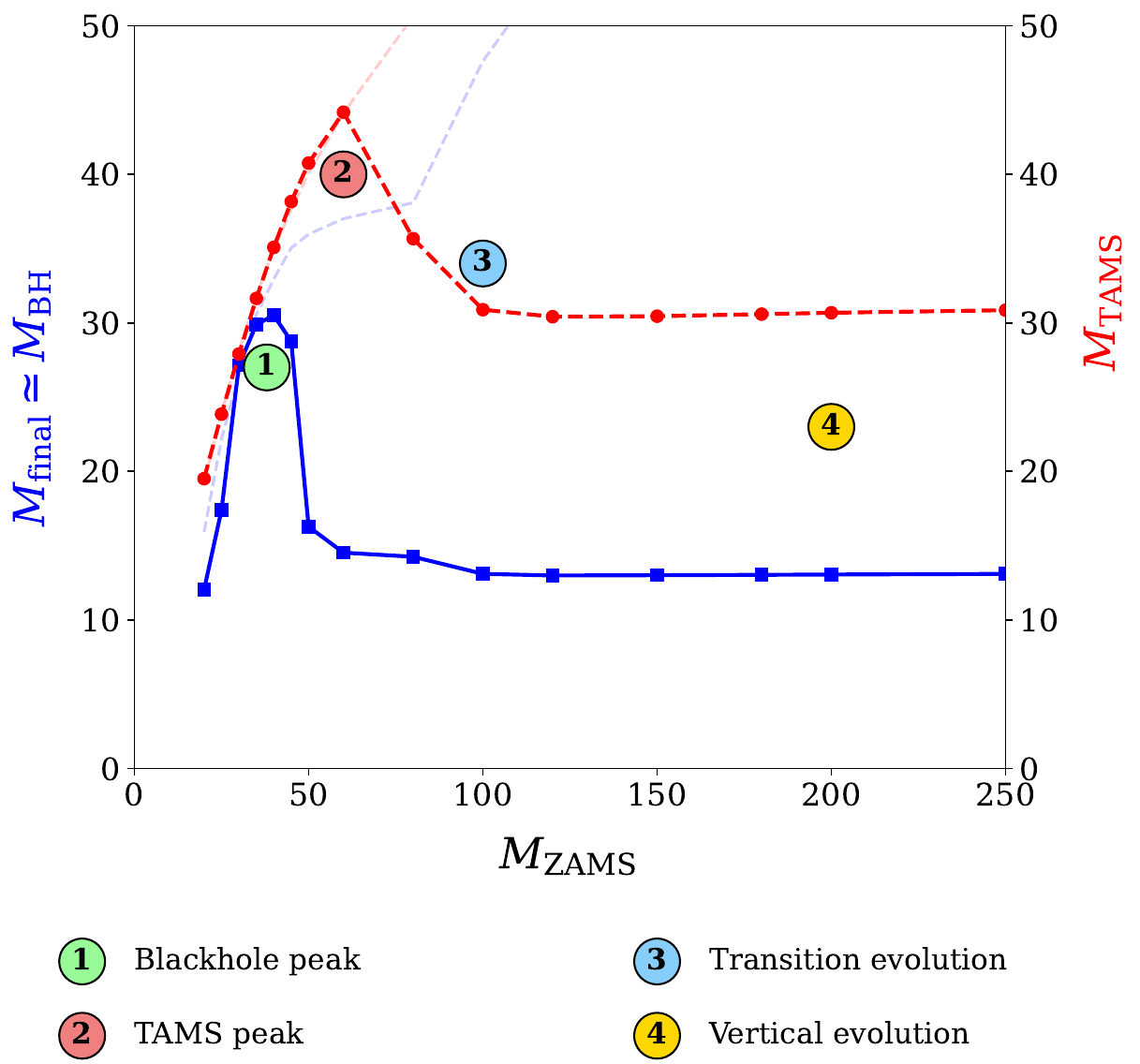}
   \caption{Initial-final mass relation for the mass range of 20, 25, 30, 35, 40, 45, 50, 60, 80, 100, 120, 150, 180, 200, and 250\msun\ initial ZAMS masses. For the lowest ZAMS masses, the final compact object could be a neutron star instead of a BH. The dashed red and blue lines give the TAMS and BH relation {\it when we only consider canonical low $\Gamma$ \cite{vink2001} winds}. The highest final BH masses are not found at the highest initial ZAMS masses, but peak around $\simeq$40\msun\ ZAMS masses, giving $\sim$30\msun\ final BH masses instead.
              \label{fig:maxBH}}
    \end{figure}

    In order to not just provide final masses from our models, but to gain insight into their qualitative behaviour, 
we first evaluated the mass at the terminal main sequence (TAMS) and then considered the temperature (\teff) during core He-burning. 

The ultimate property of interest is the final mass, which we assumed to be equal to the BH mass \citep{Fryer12,fernandez18}, but see \cite{Burrows23}. 
The initial-final mass results over the entire mass range are shown Fig.\,\ref{fig:maxBH}. 

The highest BH mass (the BH peak at 1 in Fig.\,2) is not found for the highest ZAMS masses, as might have been expected, but instead, the maximum BH mass (of 30\,\msun) is found for the ZAMS mass range between 35 and 45\msun. The 35\msun\ model looses as little as 4\msun\ during core H-burning and an additional 1\msun\ during core He-burning.

The reason for the seemingly counter-intuitive behaviour of the constant BH mass tail (4 in Fig.\,2) is that we included a $\Gamma$-dependent VMS mass loss, which takes hold above the 80-100\Msun\ transition ZAMS mass, and which causes the inflection point in the TAMS mass behaviour at point 3 in Fig\,2, although it should be noted that the switch to high-$\Gamma$ winds does not automatically imply very strong mass loss, but that this takes some time to build up. For the highest-mass stars, the mass-loss rates are sufficiently high to not only remove the H envelope, but to even remove mass from the stellar core. The BH masses from these stripped WR stars are 10-15\msun\ \citep{Higg21} at most at \Zsun.
For the mass range below 50\msun, core He-burning does not take place in the WR phase, but during the (blue, yellow, or red) supergiant phase.

The post-He-burning lifetime comprises just 1\% of the stellar evolution, so this mass loss only becomes relevant if the mass-loss rate after core He burning were to be 100 times more powerful than during the main sequence, or 10 times more powerful than during core He-burning phases (see Eq.\,2). As we cannot exclude such enhanced mass loss phases, for instance via episodic mass loss, or even eruptions, we employ an alternative upper supergiant limit approach below. 

\section{Discussion of the maximum BH mass}

Figure\,2 was computed using one particular set of physical input choices (e.g, an intermediate amount of overshooting, $\alpha_{\rm ov} \sim 0.3$; \citep{brott11}), no rotation, and $\Gamma$-dependent mass-loss rates.
However, there are many uncertainties in both mass loss and mixing (in particular, $\alpha_{\rm ov}$ and MLT/MLT$++$), and these decide whether a star stays in the blue (for negligible $\alpha_{\rm ov}$ and MLT$++$) or becomes red. The maximum BH mass in our models is reached when the star is kept in the blue. If no such low mixing were to occur in nature and the star were to move over to the red, however, strong RSG mass loss may potentially occur. In this case, this would imply that the BH mass decreases, it will not increase.  

Given that stellar evolution is non-linear and has many interdependent physical quantities related to both mixing and mass loss, we require an alternative strategy to infer the maximum BH mass from the supergiant phase.
In other words, we need to consider 
the maximum stellar mass that
stays below 
the observed maximum luminosity limit (HD-limit) for the Milky Way. According to 
\cite{Macdonald22}, this is found to be at $\log$ ($L/\Lsun)_{\rm HD} = 5.5$.

We experimented using negligibly low amounts of additional mixing ($\alpha_{\rm ov} \simeq0$) and non-boosted standard \cite{vink2001} mass-loss rates, which both helped to keep 
the stars blue by preventing potential high mass-loss episodes as an RSG. 
Basically, our pragmatic approach of considering BH progenitor stars below the HD limit simulates what occurs for stars when they approach the Eddington limit, and where they may be expected to undergo LBV-type mass loss, or more general $\Gamma$-dependent mass loss, independent of its underlying effective temperature \citep{VS23}.
As the $\alpha_{\rm ov}$ parameter cannot become negative, 30\msun\ is indeed the {\it maximum} BH mass for the upper RSG luminosity of \citep{Davies20,Macdonald22}.
We also tested the same models including rotation, and while moderate rotation
($\Omega$ of 40\% critical) increased the initial ZAMS luminosity somewhat, its further evolution was almost identical to those shown in Fig.\,1. 

To summarise, the amounts of mixing in terms of $\alpha_{\rm ov}$ and $\Omega$ are irrelevant for setting numbers 3 and 4 in Fig.\,2., and whilst they do affect the numerical values of 1 and 2, the shape of these additional features 1 and 2 is also invariant to mixing.
 
The low $\Gamma$ \cite{vink2001} mass-loss recipe includes a bi-stability jump, where the mass-loss rate is predicted to increase by a factor of about 5, but which is empirically still elusive \citep{Crow06,MP08}. 
On balance, we may have over- rather than underestimated the mass-loss rates in this mass regime \citep{petrov16,krticka17,sundqvist19}. 
The overall mass loss is already modest. It drops from 35\msun\ on the ZAMS to just 31\msun\ on the TAMS and 30\,\msun\ by the end of core He-burning; that is, 5\,\msun\ were removed from its initial value. We therefore consider the uncertainty of stationary wind mass-loss rates on the maximum BH mass to be $\sim$ 5 \msun.
In order to provide an exact error bar on our maximum BH mass, we would need to perform a N-dimensional parameter search, which is unfeasible with current machines, as the number of physical parameters of mixing and mass loss is manifold. 

The larger uncertainty in our maximum BH assessment is related to the empirical HD limit, which could indirectly account for episodic LBV-type mass loss \citep{Grassitelli21}. We have employed the \cite{Macdonald22} value of $\log(L/\Lsun)_{\rm HD} = 5.5$. Had we used the original higher value of $\log(L/\Lsun)_{\rm HD} = 5.8$ \citep{Hump79} that was in use over the last four decades, our maximum BH mass could be a high as 40\msun\ \citep{higgins19,higgins2020}. 
On the lower maximum BH side, if all uncertainties in both mixing and mass loss were to conspire to very high mass-loss rates inside the canonical O star 20-60\,\msun\ range, the maximum BH mass would simply be given by this lower limit of $\sim$ 20\,\Msun\, where mass-loss rates are negligibly low and envelopes are not removed.
When all relevant factors we discussed above are taken into account, the realistic maximum BH mass for $\Zsun$ is $M_{\rm final} = 30 \pm 10$ \msun.

\section{Final words}

The analysis performed above implies that it would be highly unlikely that for instance a 70\msun\ BH would be discovered at solar $Z$. This is not to say that a 70\msun\ BH could not exist in the Milky Way, but it should have formed at an earlier epoch when its $Z$ was still below $\sim$10\% of \Zsun\ \citep{Vink21}. As shown in Figure\,3 of \citep{Vink21} we anticipated the maximum BH mass at solar metallicity to be set by wind mass loss. As shown by Eq.(2) above, the total mass lost in stationary winds depends on both the temperature-dependent mass-loss rate and on the evolutionary duration ($\Delta t$). For this reason, we considered both the effects of wind mass loss and interior mixing, but we concluded that the maximum BH mass for \zsun\ is set by the HD limit. Therefore, the error bars on our maximum BH mass could potentially be refined by future analyses of the HD limit at \zsun.

At face value, our results for the maximum BH mass at $\zsun$ appear to be similar to those of \cite{Bavera23} and \cite{Romag24}, but there are fundamental qualitative, as explained in Fig.\,2, as well as quantitative differences. Firstly, while \cite{Romag24} included a boosted $\Gamma$-dependent mass loss, they employed a mathematical $\Gamma$ transition point, which we do not subscribe to (see the discussion in Sect.\,2.2). This appears to be the main reason for the divergent results at high $M_{\rm ZAMS}$ and the prime reason for the constant BH mass tail we find (4 in Fig\,.2). Instead, \cite{Romag24} reported an increasing BH mass with ever-increasing ZAMS mass, which is similar to our fiducial models that have no mass-loss boost (the dashed lines in Fig\,2.)

Secondly, it is interesting that the shape of our initial mass final-mass relation (Fig.\,2) shows qualitative resemblances to the shapes reported by \cite{Bavera23}, although they employed a simple constant LBV-like mass loss while we employed an exponentially increasing $\Gamma$-dependent mass loss. While the peak of the mass function is reached at a similar BH mass, the initial mass to which this BH mass corresponds is twice as high in \cite{Bavera23}. We attribute this to our strategy of employing the empirical HD limit. 

We would like to raise one potential caveat to our analysis. For our results on the maximum BH mass at solar metallicity, we used enhanced mass-loss rates for VMS, which led to our conclusion that VMS do not yield the maximum BH mass at solar $Z$. If we have over-estimated mass-loss rates for VMS using the \cite{Sabh22} implementation, then our analysis would be notably affected. The reason we nonetheless should have confidence in the high mass-loss rate implementation for VMS is that this solution alone naturally explains the almost constant effective temperatures of VMS at various metallicities, whereas lower mass-loss rates would need to be tweaked for each and every VMS mass bin value to either avoid redward inflation resulting from mass-loss rates that are too low or a blueward evolution due to high mass loss (see \cite{Sabh22} for a detailed discussion).

We would like to conclude by reflecting on the fact that there are still many uncertainties in both interior mixing and stellar wind mass loss,
but using key observational and theoretical constraints, we have shown that it is possible to provide a realistic maximum BH mass. 
If a BH were on day discovered at solar $Z$ that is substantially higher than 30-40\msun, we would need to seriously reconsider mass-loss rate implementations for VMS because we have exhausted all other relevant physical uncertainties in detailed stellar evolution modelling.

\begin{acknowledgements}
The authors acknowledge MESA authors and developers for their
continued revisions and public accessibility of the code. 
The authors would also like to thank the anonymous referee for their constructive comments which helped improve the presentation of the results.
We are supported by STFC
(Science and Technology Facilities Council) funding under grant
number ST/V000233/1. 
\end{acknowledgements}

\bibliographystyle{aa}
\bibliography{references}

%\begin{thebibliography}{}
%\end{thebibliography}

\end{document}